\documentclass{aa}
\usepackage[varg]{txfonts}
\usepackage{graphicx}
\usepackage{hyperref}
\usepackage{natbib}

\usepackage{bm}

\bibpunct{(}{)}{;}{a}{}{,}

\begin{document}    

\title
{Bayesian estimate of the superfluid moments of inertia from the 2016 glitch in the Vela pulsar}

\author{  
        A.~Montoli\inst{\ref{inst1}, \ref{inst2}}\thanks{alessandro.montoli@unimi.it}    
        \and    
        M.~Antonelli\inst{\ref{inst3}}\thanks{mantonelli@camk.edu.pl}  
        \and
        F.~Magistrelli\inst{\ref{inst1}}
        \and
        P.~M.~Pizzochero\inst{\ref{inst1}, \ref{inst2}}  
}    
    
\institute{    
        Dipartimento di Fisica, Universit\`a degli Studi di Milano, Via Celoria 16, 20133 Milano, Italy\label{inst1}    
        \and                                                                                               
        Istituto Nazionale di Fisica Nucleare, sezione di Milano, Via Celoria 16, 20133 Milano, Italy\label{inst2}    
        \and    
        Nicolaus Copernicus Astronomical Center of the Polish Academy of Sciences, ul. Bartycka 18, 00-716 Warszawa, Poland\label{inst3}    
}    

\abstract
{The observation of the first pulse-to-pulse glitch in the Vela pulsar opens a new window on the internal dynamics of neutron stars as it allows us for testing models in the circumstances of the first moments of a glitch. Several works in the literature have already considered the observational and physical parameters of the star by employing a minimal model with three rigidly rotating components.}
{We analytically study the minimal three-component model for pulsar glitches by solving it with generic initial conditions. The purpose is to use this solution to fit the data of the 2016 Vela glitch by employing a Bayesian approach and to obtain a probability distribution for the physical parameters of the model and for observational parameters, such as the glitch rise time and the relaxation timescale.
}
{The fit is achieved through Bayesian inference. Due to the presence of an increase in the timing residuals near the glitch time, an extra magnetospheric component was added to the three-component model to deal with this phenomenon. A physically reasonable, non-informative prior was set on the different parameters of the model, so that the posterior distribution could be compared with state-of-the-art information obtained from microphysical calculations. By considering a model with a tightened prior on the moment of inertia fractions and by comparing it with the original model by means of Bayesian model selection, we studied the possibility of a crust-limited superfluid reservoir.
}
{We obtained the posterior distribution for the moment of inertia fractions of the superfluid components, the coupling parameters, and the initial velocity lags between the components. An analysis of the inferred posterior also confirmed the presence of an overshoot in that glitch and set an upper limit of $\sim 6\,$s on the glitch rise timescale. 
The comparison between the two models with different priors on the moment of inertia fractions appears to indicate a need for a core participation in the glitch phenomenon, regardless of the uncertain strength of the entrainment coupling.
}{}

\keywords{dense matter - stars:neutron - pulsars:general - pulsars:individual:PSR J0835-4510}

\maketitle


\section{Introduction}
\label{sec:intro}
%
%
One of the timing anomalies observed in the regular emission from radio pulsars are glitches, which are sudden accelerations of the rotation of a pulsar followed by a slow relaxation towards a post-glitch phase of slow and smooth spin down \citep[see e.g. ][]{lyne2000,espinoza+2011}. 
These events are quite rare and capturing an observation of a glitch in the act requires continuous monitoring.

The modelling of pulsar glitches requires at least two different components in the star 
\citep{baym+1969}: a normal component, which is coupled on short timescales to the magnetosphere, and a superfluid component, which stores angular momentum by pinning it to impurities in the crust \citep{andersonitoh1975}  or to fluxtubes in the core of the star \citep{alpar2017JApA}. 
This reservoir occasionally releases angular momentum to the observable normal component of the star, giving rise to the glitch, even though the matter of exactly what triggers the glitch itself is still under debate \citep{haskellmelatos2015}.
On the other hand, the location of the superfluid reservoir is also unknown. 
For years, it has been assumed to be located in the crust of the star due to the fact that about the $1.7\%$ of the Vela spin-down is reversed, on average, during a glitch -- a proportion that is close to what it is theoretically expected for the moment of inertia fraction carried by the unbound neutrons in the crust \citep{dattaalpar1993,link+1999}. 
However, recent works have shown that it is not possible to account for the large glitches of the Vela pulsar if we limit the superfluid reservoir  to just the dripped neutrons in the inner crust \citep{andersson+2012, chamel2013, carreau2019} and that at least a small shell of the outer core of the star has to be taken into account \citep{ho+2015,montoli2020eos}. 
These results are based on a statistical parameter of a glitching pulsar, also known as activity, which quantifies the mean spin-up rate of the star due to glitches \citep[e.g. ][]{fuentes17}, as well as on a microphysical parameter, the entrainment coupling, which is responsible for a non-dissipative interaction between the two components of the star (see \citealt{haskellsedrakian2017,chamel_super} for recent reviews). 
The values of the entrainment parameter in the crust and the scale of its effects are, however,  still open to discussion \citep{martinurban2016, Watanabe2017,chamel2017JLTP,sauls+2020}.

Up until now, most of the information about pulsar glitches was obtained through the analysis of the frequency and sizes of glitches, which, together with theoretical modelling, have been used to constrain the structure of neutron stars \citep{ho+2015,delsate2016,pizzochero+2017}.
More recently, a glitch of the Vela pulsar in December 2016 has been observed with unprecedented precision, making it possible to detect every single pulsation during the glitch \citep{palfreyman+2018}. 
This kind of measurement has opened new possibilities for contrasting our theoretical understanding of the glitch phenomenon with timing data \citep{graber+2018, ashton+2019, pizzochero+2019,guerciVela2016}.

To extract useful information from the data reported by \citet{palfreyman+2018}, \citet{ashton+2019} performed a Bayesian fit  to the time of arrival (TOA) of the single pulses with an `agnostic' timing solution containing few parameters that set the glitch amplitude and the typical timescales of the process. 
In this way, it has been possible to put an upper limit to the spin-up rise time of $\sim 12\,$s, lowering the early $\sim40\,$s boundary \citep{dodson+2002, dodson+2007}, and to confirm the presence of an overshoot - in other words, the acceleration of the rotation of the star up to velocities larger than that of steady state equilibrium - in the data.

In \citet{pizzochero+2019}, the agnostic timing solution used by \citet{ashton+2019} has been written in terms of the parameters of a minimal three-component model - two superfluid and one normal - for the pulsar rotation.
These parameters are linked to structural and rotational properties of the glitching star, such as the fractions of the moment of inertia of the superfluid components, the initial lag between the two superfluid components and the normal one, and the coupling parameters between them. 

The aim of this paper is to extend the formalism of the theoretical discussion of three-component models presented in \citet{pizzochero+2019} - hereafter Paper I - and to refine the statistical procedure behind the least mean squares fit made there by employing a fully Bayesian approach. The difference with respect to the analysis of \citet{ashton+2019} is that here, the underlying fit model is not agnostic but it has a clear interpretation in terms of the minimal three-component model used to describe the underlying pulsar dynamics: this allows us to fit not only the glitch timescales and the glitch amplitude, but to also infer  some structural properties of the neutron star.

In contrast to  what was done in Paper I, here we employ the whole dataset of \citet{palfreyman+2018}.
As noted in previous works \citep{palfreyman+2018, ashton+2019}, an increase of the timing residuals has been detected in the vicinity of the glitch time.
This can be interpreted as the signature of a magnetospheric change accompanying the glitch \citep{palfreyman+2018}.
Since a three-component model is equipped to take this kind
of phenomena into account, additional modelling is needed.
In \citet{ashton+2019}, an extra phenomenological term was introduced to deal with this decrease in the rotational frequency.
In this work, we  extend the three-component model, speculating on an instantaneous decoupling and recoupling of the magnetosphere to the crust.


This paper is outlined as follows: in Section~\ref{sec:model}, we present the three-component model, extended with respect to the one used in Paper I, along with theoretical considerations on the glitch overshoot occurrence.
In Section~\ref{sec:analysis} ,  we first present some considerations about the data set of \citet{palfreyman+2018} and then  describe the statistical modelling behind the fit we performed.
The results of the Bayesian fit are described in Section \ref{sec:results} and \ref{sec:physical}. 
The appendices are devoted to technical aspects of glitch models with three rigid components: 
in Appendix \ref{sec:solution}, we derive the general solution of the three component model (i.e. we extend the solution in Paper I by allowing for general initial conditions). In  Appendix \ref{sec:constr}, the constraint on the moment of inertia of the superfluid component found by \cite{sourie2020mnasL} is derived in the present, more general, setting. In Appendix \ref{app_entr}, we show how it is possible to take into account the entrainment coupling in glitch models with several rigid components.


\section{Model}
\label{sec:model}

Following the seminal idea of \cite{baym+1969}, we model a glitch by formally dividing a spinning neutron star into several rigidly rotating components that can exchange angular momentum. 
The minimal  model we adopt here consists of two superfluid components - corresponding to different, non-overlapping regions of the star where neutron superfluidity is expected - and a normal component that extends over the whole stellar interior. 
The normal component, labelled with the $p$ subscript, is usually believed to be coupled with the magnetic field of the star and, thus, observable from Earth, while the two superfluid components (labelled 1 and 2) act as reservoirs for angular momentum and their rotation cannot be tracked from Earth. 

Following Paper I, we assume that the two superfluid components do not interact directly between themselves, but they interact only with the normal component. 
The strength of this interaction is set by two phenomenological coupling parameters, $b_{1}$ and $b_2$. 
Finally, all these three components lose angular momentum with a constant rate given by the electromagnetic braking torque. 
Therefore, the system of equations for the evolution of these three components is the natural three-component extension of the dynamical system introduced by  \cite{baym+1969}, namely,
%
\begin{align}
\begin{split}
        & x_p \dot{\Omega}_p + x_1 \dot{\Omega}_1 + x_2 \dot{\Omega}_2 = - |\dot{\Omega}_\infty|\\
        & \dot{\Omega}_1 = -b_1 \left( \Omega_1 - \Omega_p \right)\\
        & \dot{\Omega}_2 = -b_2 \left( \Omega_2 - \Omega_p \right),
\end{split}
\label{eq:3c}
\end{align}
where $x_j$ with $j = \{1,\ 2,\ p\}$ is the fraction of moment of inertia of the $j$-th component with respect to the total moment of inertia, and $|\dot{\Omega}_\infty|$ is the steady state spin down. The partial moments of inertia must sum up to the total one, so that we impose
\begin{equation}
\label{uno}
    x_p + x_1 + x_2 = 1 \, .
\end{equation}
The system in \eqref{eq:3c} is valid for $t \geq 0$, where we have set $t=0$ as the time at which the glitch is triggered. 
Prior to the glitch moment, the values of $b_{1,2}$ could have a different value, for example, they may be assumed to be equal to zero if the two superfluid components are perfectly pinned at $t<0$, but their actual pre-glitch value is not important for our scope. 
Since $b_{1,2}$ set the post-trigger creep rate of vortex lines \citep{Alpar84a}, what is important in the present analysis is that their values remain almost constant during the glitch spin-up phase and the first moments of the relaxation \citep[see e.g.][for models where these mutual friction coefficients are functions of the velocity lag between the components]{celora2020arXiv}. 
Hence, a limitation of the model would be to drop the still poorly understood problem of the post-glitch repinning process, during which the creep rate is expected to decrease as the velocity lag between the superfluid and the normal component becomes smaller and smaller \citep{sedrakian_repinning1995,Haskellhop}.



The system in~\eqref{eq:3c} is solved in Appendix \ref{sec:solution} by making a change of variables: instead of $\Omega_i$, it is more convenient to use the angular velocity lags $\Omega_{ip}=\Omega_i-\Omega_p$ between the $i$-th superfluid component and the normal one. 
The first part of Equation~\eqref{eq:3c} can be integrated to obtain the angular velocity residue: 
\begin{equation}
        \Delta \Omega_p(t) \equiv \Omega_p(t) - \Omega^0_p + t |\dot{\Omega}_\infty|,
        \label{eq:omegap}
\end{equation}
where $\Omega_p^0 = \Omega_p(0)$ is the angular velocity of the normal component just before the glitch starts. 
The dynamics of the two lags, $\Omega_{ip}$ , must satisfy the other two equations of the system, that can be solved for arbitrary initial conditions $\Omega_{ip}(0) = \Omega_{ip}^0$; see Appendix~\ref{sec:solution} (on the contrary, the solution presented in Paper I is a particular one since it is valid only for a particular subset of initial conditions for the lags). 
The general form of the solution  has the form (cf. with Paper I and \citealt{ashton+2019}):
\begin{equation}
\Delta \Omega_p(t) 
\, = \, 
\Delta\Omega_p^\infty \left[ 1- \omega \, e^{-t \lambda_+} - (1-\omega)\,  e^{-t \lambda_-}  \right] ,
\label{eq:dOmegap}
\end{equation}
where $\omega$, $\lambda_{\pm}>0$ and $\Delta \Omega_p^\infty$ are time-independent functions of the parameters in the system \eqref{eq:3c}, defined in \eqref{omega}, \eqref{werty}, \eqref{qwer}, and \eqref{eq:lambdas}.  

Since $0<\lambda_-<\lambda_+$, we have that $\Delta \Omega_p^\infty$ is the asymptotic value of the glitch amplitude (depending on the initial conditions it could be either positive or negative).
Moreover, $\Delta \Omega_p^\infty$ can be thought to represent the glitch jump that could be extracted from the analysis of post-glitch timing data when the glitch is not observed in the act (see e.g. Figure 11 in \citealt{antonellipizzochero2017}). 

As already noted in previous works, the angular velocity of the observable component shows an evolution with two different timescales, one given by $1/\lambda_+$ and a longer one given by $1/\lambda_-$. 
In fact, Equation \eqref{eq:dOmegap} has the same functional form of the agnostic model used to fit the Vela 2016 glitch  by \cite{ashton+2019}; the difference here is that we make an exact connection between the `solution' parameters in Equation \eqref{eq:dOmegap} and the `structural' parameters in Equation \eqref{eq:3c}, which have a physical interpretation.

Next we study the conditions under which an overshoot of the normal component can be produced, a situation that can never occur in a model with only two rigid components and a constant coupling parameter.
We note, however, that it is possible to obtain an overshoot with a two-component model of the kind pionereed by \cite{alpar_first_creep_1981}, where the superfluid component can develop non-uniform rotation 
\citep[see e.g.][]{Alpar84a,larson_link_2002,haskell+2012,antonellipizzochero2017,graber+2018} due to the fact that the coupling with the normal component - which depends on the non-uniform lag itself and on stratification - may not be constant in both space and time. This is not surprising as a fluid model has infinite degrees of freedom that can react on different timescales -- and not just two (i.e. $\Omega_1$ and $\Omega_2$), as in the present minimal model.

The overshoot is realised if there exists a certain time, $t_{\rm max}>0,$ such that $\Delta\dot{\Omega}_p=0$ and $\Delta\ddot{\Omega}_p<0$. 
The first derivative of  Equation \eqref{eq:dOmegap} gives: 
\begin{equation}
        t_{\rm max} \, = \, \frac{1}{\lambda_+ - \lambda_-} \, 
\left[ 
\log \left( \frac{\lambda_+}{\lambda_-} \right) 
+
\log \left( \frac{ \omega}{\omega-1} \right)
\right]\, ,
\label{eq:tmax}
\end{equation}
which needs to be positive. Since $\lambda_+ > \lambda_- > 0$ (see Equation \ref{eq:eigenvalues}), we have that  $t_{\rm max}$ is a real number when $\omega <0$ or $\omega > 1$. 
The additional condition, $\Delta \ddot{\Omega}_p(t_{\rm max}) < 0,$ requires that $\omega > 0$. 
Therefore, the overshoot occurs for $\omega > 1$, which also guarantees that $t_{\rm max} > 0$. In particular, we have a very delayed overshoot if $\omega \rightarrow 1^+$, while for $\omega \rightarrow +\infty$ we find a lower bound to the duration of the spin-up phase, namely, 
\begin{equation}
t_{\rm max} \, > \, 
\frac{ \log \left( \lambda_+/\lambda_- \right) }{ \lambda_+ -\lambda_- } \, . 
\end{equation}
The condition for an overshoot can be translated in terms of time residuals with respect to the steady spin-down evolution, which are given by  \citep[cf. ][]{graber+2018}
\begin{equation}
        r_p(t) = - \frac{1}{\Omega_p^0} \int_0^t \Delta \Omega_p(t') \mathrm{d}t' \, .
        \label{eq:rG}
\end{equation}
For an overshooting glitch, $t_{\rm max}$ corresponds to a flex point, after which $\ddot{r}_p(t)$ is positive. 
On the other hand, in a non-overshooting glitch there is no flex point and  $r_p(t)$ is always concave down. 

We note that all the equations above are symmetric under the exchange of the label 1 with 2.
Therefore, to break this degeneracy and physically distinguish one superfluid component from the other, we impose that the superfluid component 2 is the one with the biggest initial lag. 
This may be due to a stronger pinning in the region of  component 2 or simply because it happened that the glitch initiated in this condition (the initial conditions are unknown and depend on the past history of the system). 
Hence, the superfluid component 1 is, by definition, the one with a smaller initial lag, 
\begin{equation}
    \Omega_{1p}^0 \, < \, \Omega_{2p}^0 \, .  
    \label{break12}
\end{equation}


\section{Analysis of the 2016 Vela glitch}
\label{sec:analysis}

Following a   Bayesian approach,  we  find the posterior probability distribution for the phenomenological parameters of the model in Equation \eqref{eq:3c}.

\subsection{Magnetospheric change and data set}
\label{subsec:data}

The data made available by \cite{palfreyman+2018} span a 4200s time window, with the glitch time positioned roughly at the centre of the dataset. 
The authors calculate a first estimate of the glitch date, set at $t_g^P = 57734.4849906$ MJD. 
Moreover, they identify some peculiarities during the glitch: at a time, $t_1 =t_g^P-1.5$s, an increase of the residuals is detected. 
This kind of behaviour can be linked to an effective slow-down of the star before the actual glitch \citep{ashton+2019} or to a magnetospheric change \citep{palfreyman+2018} that could cause a delay on the emission of the pulsations of the star, perhaps even due to a starquake \citep{bransgrove+2020}.  
Of course, this phenomenon cannot be described using the model presented in Section~\ref{sec:model}, thus, further modelling is necessary  to fit the timing data. 
To do so, we assume that the magnetosphere instantaneously decouples and recouples from the rotation of the crust of the star, lagging behind the actual angular velocity of the charged component. 
This amounts to introducing a fourth component with negligible inertia  (the magnetosphere) that is always locked to the $p$-component apart for an instantaneous jump at $t=\Delta t_M$, namely, 
\begin{equation}
        \Omega_M(t) = \Omega_p(t) - \Omega_p^0 \Delta r_0 \delta(t -  \Delta t_M),
        \label{eq:orM}
\end{equation}
where $\Delta r_0$ and $ \Delta t_M$ are additional phenomenological parameters that  have to be fitted  together with $x_i$, $b_i$, and the initial lags, $\Omega_{ip}$. 
Due to its instantaneous behaviour, Equation \eqref{eq:orM} is  non-physical, but it provides a simple mathematical form for this magnetospheric slip, which is suggested by the data; its impulsive character is a crude simplification of a complex dynamical problem. 
Hence, the modelling in \eqref{eq:orM} represents the minimal choice  to extend the system \eqref{eq:3c} to take into account this additional piece of physics that is present in the data of \cite{palfreyman+2018}.

The residual function of the `observable component' (that is now the magnetosphere) takes the form:
\begin{equation}
        r_M(t) = r_p(t)  \vartheta(t) + \Delta r_0 \vartheta(t - \Delta t_M),
        \label{eq:rM}
\end{equation}
where we extended the function, $r_p(t),$ to pre-glitch times, $t<0,$ by means of the Heaviside step function $\vartheta$.  The sign of $\Delta t_M$ is unknown and can be either negative (the magnetospheric change happened before the glitch) or positive (the magnetospheric change follows the glitch trigger).
Finally, the data provided by \citet{palfreyman+2018} are lacking with regard to the uncertainty on the single measure of the residual. We estimate it from the standard deviation of all the data before $t_1$, as it is  quite certain that before that time, the star had not yet undergone the glitch. 
In this way, we find $\sigma=0.25\,$ms and we assume this value to be valid also for the post-glitch measurements.

\subsection{Bayesian modelling}
\label{subsec:bayes}

We now describe the statistical modelling used to obtain a probability distribution for the parameters involved in the model. 
From Equation~\eqref{eq:rG} we have up to a maximum of six parameters: the two coupling parameters, $b_{1,2}$; the two moment of inertia fractions, $x_{1,2}$ ; and the two initial lags $\Omega_{1,2\, p}^0$. 
The number of these parameters can be reduced by assuming the initial lag for the component 1 to be that of the steady state, so that the model can be simplified  to the one discussed in Paper I. 
Here, we keep the discussion as general as possible, thus, keeping $\Omega_{1p}^0$ as a free parameter to be fitted.

The residuals of Equation~\eqref{eq:rG} have to be considered with respect to the glitch date, $t_g$, which is itself a parameter of the model. 
Moreover, the magnetospheric slip defined in Equation~\eqref{eq:rM} has to be included in the model as well. 
In other words, the residual function, $r(t),$ which describes all the pre-glitch and post-glitch data is:\ 
\begin{equation}
        r(t) = r_M(t - t_g) = r_p(t- t_g)  \vartheta(t-t_g) + \Delta r_0 \vartheta(t-t_M) \, ,
        \label{eq:residuals}
\end{equation}
where $t_M = t_g+\Delta t_M$ is the date of the magnetospheric slip.
In the following, the estimate of these two date parameters, $t_g$ and $t_M$,  is given with respect to the glitch date $t_g^P$ calculated in the analysis of \citet{palfreyman+2018}.

We collectively note all the nine  parameters of the model as:\  
\begin{equation}
\label{otto}
 \mathcal{P} \, = \, 
\{  \,x_1, \,x_2, \,b_1, \,b_2, \,\Omega_{1p}^0, \,\Omega_{2p}^0,  \,\Delta r_0, \,t_g, \,t_M \, \}    
\end{equation}
The probability distribution for these parameters can be obtained as the posterior distribution of a Bayesian inference \citep{mackay_book},
\begin{equation}
        P(\mathcal{P}\, |\, \mathcal{D}) = \frac{P(\mathcal{D}\, |\, \mathcal{P})\, P(\mathcal{P})}{P(\mathcal{D})}\, ,
        \label{eq:bayes}
\end{equation}
where the functions  $P(\mathcal{D}\, |\, \mathcal{P})$, $P(\mathcal{P})$ and $P(\mathcal{D})$ are the likelihood, the prior and the evidence, respectively and
\begin{equation}
\mathcal{D} \, = \, \{  \, (\,  t_i \, , \, r_i \, ) \, \}_{ \, i \, \in \, \text{data}}
\end{equation}
represents the data used for the fit, that is, the time of arrival of the pulses, $t_i$ , and the measured residual, $r_i$ , with respect to the model of a uniform spin-down. 

Assuming that the measurement for a single pulsation is independent of the measurements of the others, we  write the likelihood as (see also \citealt{ashton+2019}):
\begin{equation}
        P(\mathcal{D}\, |\, \mathcal{P},\, \sigma) = \prod_i \frac{1}{\sqrt{2\pi\sigma^2}}\exp\left(-\frac{(r(t_i) - r_i)^2}{2\sigma^2}\right)\, ,
        \label{eq:likelihood}
\end{equation}
where $\sigma$ is the uncertainty on the single measure as calculated in Section~\ref{subsec:data}. 
By writing the likelihood in this way, however, we made a further simplification: here, the uncertainty $\sigma$ is referred only to the time residual, $r_i$, while the same uncertainty must affect the time of arrival $t_i$ as well, as the two quantities are interdependent. 
In fact, an hypothetical variation of the time of arrival would generate the same variation in $r_i$ and vice versa. 
Thus, the correct likelihood should be a normal distribution with variance $\sigma^2$ and set diagonally 
on the $(t_i,r_i)$ space. 
Since the uncertainty on the measure of the TOAs is of the order of a fraction of ms, while the pulsations arrive on timescales of a tenth of a second, we  neglect this correction and use the distribution in Equation \eqref{eq:likelihood}.

We assume most of the variables to be independent from the others, so to factorise  the prior for the parameters $P(\mathcal{P})$ into smaller parts. 
We set the probability distribution of the moment of inertia fractions $x_i$ as a uniform distribution between 0 and 1, with the constraint that the sum is less than unity, 
\begin{equation}
        x_1,\, x_2 \sim \begin{cases} \text{Unif(0,1) Unif(0,1)}  &\text{if } x_1 + x_2 < 1 \\ 0 &\text{elsewhere} \end{cases}
        \label{eq:Px}
.\end{equation}
%
For each of the two coupling parameters $b_i$ we chose a log-uniform distribution as we do not know the order of magnitude of the coupling parameters and we would like to explore a wide range of orders of magnitude,
\begin{align}
        b_1\, [\text{s}^{-1}] &\sim \text{LogUnif($10^{-6}$, $10^{0}$)}
        \label{eq:Pb1}
        \\
        b_2\, [\text{s}^{-1}] &\sim \text{LogUnif($10^{-4}$, $10^{2}$)} \, .
        \label{eq:Pb2}
\end{align}
%
In this way, we took a first step in breaking the symmetry between the two superfluid components by setting two different (but overlapping) priors on the two coupling parameters. 
Similarly, we chose a log-uniform distribution for the prior of the initial lags $\Omega_{ip}^{0}$.
Since the two priors in \eqref{eq:Pb1} and \eqref{eq:Pb2} are largely overlapping, we definitively break this symmetry by setting a prior on the initial lags that automatically implements the condition \eqref{break12}
%
\begin{equation}
        \Omega_{1p}^0,\, \Omega_{2p}^0\, [\text{rad/s}]  \sim 
        \begin{cases} 
                \begin{split}
                \text{LogUnif($10^{-10}, 10^{-1}$)}\, \times 
                \\ \times \text{LogUnif($10^{-5}, 10^{-1}$)}\end{split} &\text{if }\, \Omega_{1p}^0 < \Omega_{2p}^0\\
                0 &\text{elsewhere} 
        \end{cases}
        \label{eq:Pom}
.\end{equation}
We determine the prior on the shift on the timing residuals given by the magnetospheric change to be as broad as possible: since the pulsation of the Vela has a frequency of $\approx 10\, $Hz, we set the prior on $\Delta r_0$ to be a uniform distribution between $-100$ ms and 100 ms. 
In this way, we cover a whole pulsation, which can be up to 0.1 seconds early or 0.1 seconds late,
\begin{equation}
        \Delta r_0\, [ \text{ms} ] \sim \text{Unif}(-100,\, 100).
        \label{eq:Pr0}
\end{equation}
Finally, we set the two priors on the two dates, $t_g$ and $t_M$ , respectively to be uniform between $-100$ s and 100 s and between $-1000$ s and 100 s with respect to the glitch date $t_g^P$ obtained by \citet{palfreyman+2018}. 
We do not set further conditions on the relation between them. 
In this way, it is, in principle, possible to understand whether the magnetospheric change preceeded  the glitch, or vice versa \textit{} \citep[see also][]{ashton+2019}:
\begin{align}
        t_g\, &[\text{s}] \sim \text{Unif(-100, 100),}
        \label{eq:Ptg}\\
        t_M\, &[\text{s}] \sim \text{Unif(-1000, 100).}
        \label{eq:Ptm}
\end{align}     
The whole prior distribution  $P(\mathcal{P})$  is the product of all these independent probability distributions, defined in Eqs.~(\ref{eq:Px}) to (\ref{eq:Ptm}): 
\begin{multline}
        P(\mathcal{P}) = P(x_1, x_2)\, P(b_1)\, P(b_2)\, P(\Omega_{1p}^0, \Omega_{2p}^0)\, \times
        \\
        \times \, P(\Delta r_0)\, P(t_g)\, P(t_M) \, .
        \label{eq:prior}
\end{multline}

\section{Results of the Bayesian fit}
\label{sec:results}

\begin{figure*}
        \includegraphics[width=\textwidth]{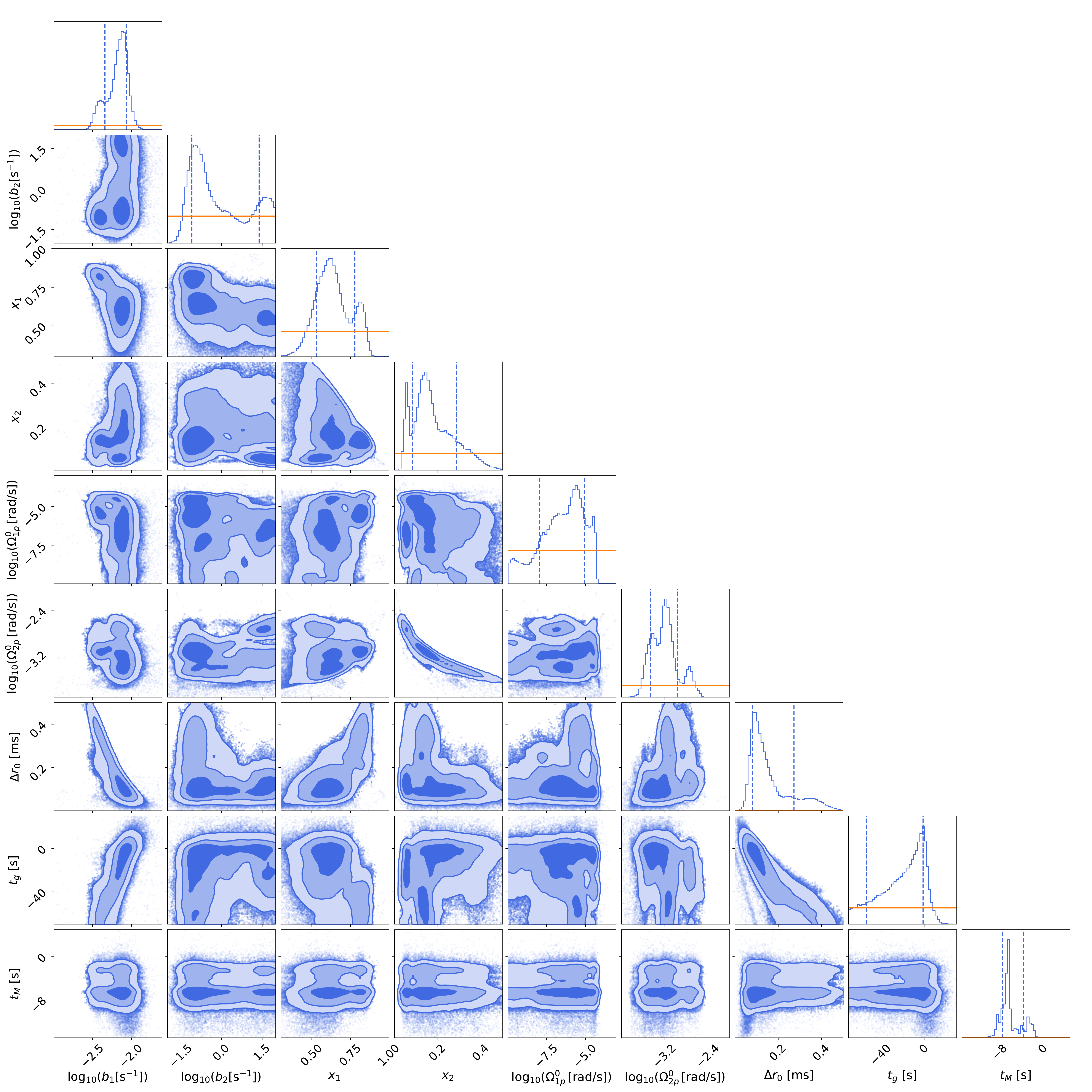}
        \caption{
        Cornerplot of the posterior distribution. 
        On the diagonal, the marginalised posterior distribution for each parameter of the model is plotted.
        The vertical lines represent the 16th and 84th percentiles of these distributions. 
        The numerical values are reported in Table~\ref{tab:9par}. 
        The prior distribution is plotted in orange as a comparison: for the jump in the residuals $\Delta r_0$ and the magnetospheric time $t_M$ this is almost invisible, due to the width of the distribution.
        The covariance plots are located off-diagonal.
	}
        \label{fig:9par}
\end{figure*}

We set the angular velocity at the time of the glitch to a value of $\Omega_0^p = 70.34$ rad/s, while for the angular velocity derivative, we use the value $|\dot{\Omega}_\infty| = -9.78 \times 10^{-11}$ rad s$^{-2}$ \citep[see e.g.][]{dodson+2002}. 
The posterior distribution for the nine parameters in \eqref{otto} has been inferred by employing the \texttt{dynesty} nested sampler \citep{dynesty}, as implemented in the \texttt{Bilby} Python package \citep{bilby}. 
The results for these nine parameters are shown in Figure~\ref{fig:9par}, with the 16th, 50th, and 84th percentiles for each variable reported in Table~\ref{tab:9par}.

\begin{table}
        \centering
        \begin{tabular}{cccc}
                Variable & 16th percentile & Median & 84th percentile\\
                \hline
                $b_1$ & 0.004 & 0.007 & 0.009 \\
                $b_2$ & 0.08 & 0.37 & 24.64 \\
                $x_1$ & 0.53 & 0.63 & 0.78 \\
                $x_2$ & 0.08 & 0.16 & 0.29 \\
                $\Omega_{1p}^0$ & $1.06 \times 10^{-8}$ & $5.18 \times 10^{-7}$ & $8.61\times 10^{-6}$ \\
                $\Omega_{2p}^0$ & $0.3 \times 10^{-3}$ & $0.6 \times 10^{-3}$ & $1.1 \times 10^{-3}$ \\
                $\Delta r_0$ & 0.08 & 0.12 & 0.27 \\
                $t_g$ & -53.1 & -18.2 & -1.1 \\
                $t_M$ & -7.59 & -6.46 & -3.61 \\
        \end{tabular}
        \caption{16th, 50th, and 84th percentiles for the marginalised posterior for the different variables of the model. The values of $b_1$ and $b_2$ are given in units of s$^{-1}$, $\Omega_{1p}^0$ and $\Omega_{1p}^0$ are in rad/s, $\Delta r_0$ in ms, $t_g$ and $t_M$ in seconds, using the date $t_g^P$ of \citet{palfreyman+2018} as reference time origin. }
        \label{tab:9par}
\end{table}

\subsection{Magnetospheric event}

\begin{figure}
        \includegraphics[width = \columnwidth]{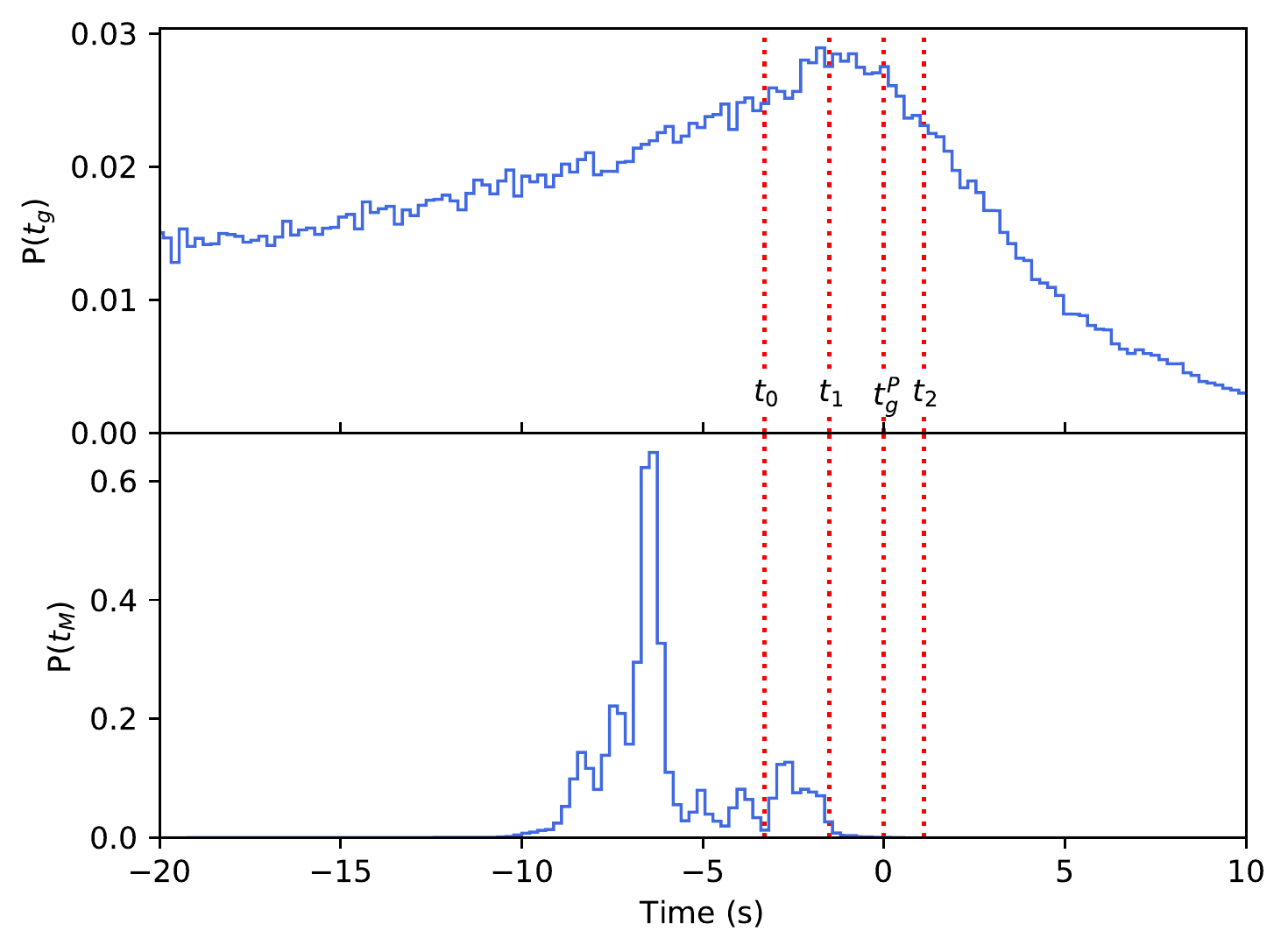}
        \caption{
        Probability distribution for the inferred glitch time $t_g$ and the time of the magnetospheric slip $t_M$. 
        For comparison, some characteristic times obtained in \citet{palfreyman+2018} are superimposed: the time of a null pulse $t_0$, the start and the end of the rise of the residuals $t_1$ and $t_2$, and the glitch time $t_g^P$ as calculated in that paper.}
        \label{fig:times}
\end{figure}

\begin{figure}
        \includegraphics[width = \columnwidth]{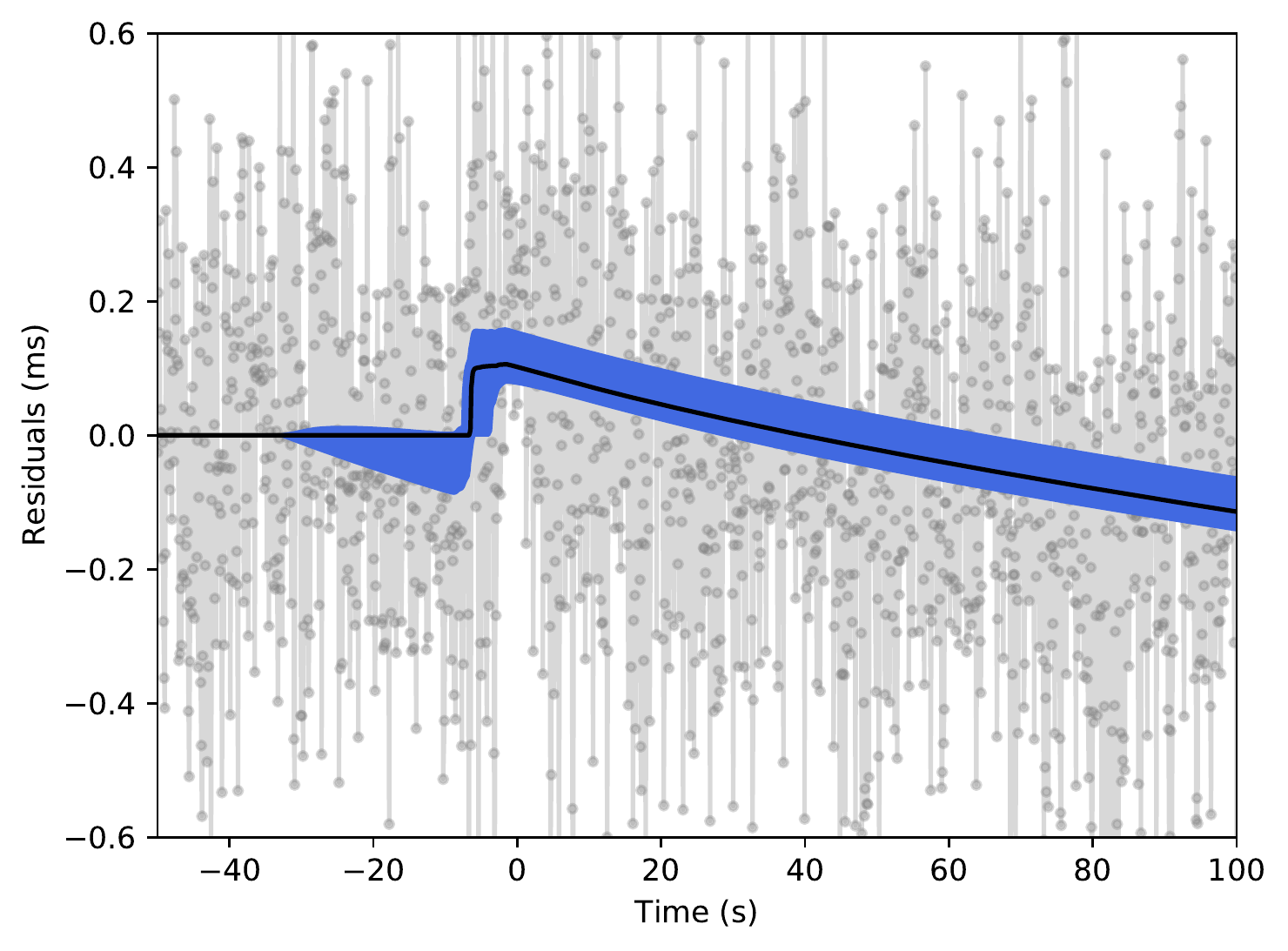}
        \caption{
        Result of the fit. We plot the data obtained by \citet{palfreyman+2018} in grey, joined by a line, and the fitted curve: for each time $t$ every $0.1$ s between $-50$ s and $100$ s, we calculate the probability distribution for $r(t)$ starting from the samples of the posterior distribution. The median of the probability distribution for the residual function $r(t)$ defined in \eqref{eq:residuals} is plotted in black, while the blue region indicates the 16th-84th percentile zone. The reference time $t = 0$ is set to be the glitch time $t_g^P$ calculated in \citet{palfreyman+2018}.
        }
        \label{fig:fit}
\end{figure}

In Figure~\ref{fig:times}, we show the two distributions for the glitch time, $t_g$, and the magnetospheric change time, $t_M$, along with some characteristic times defined in \citet{palfreyman+2018}: the authors detected a missing pulse at time $t_0$ and a persistent increase of the residuals which took place between $t_1$ and $t_2$. 
The glitch time $t_g$ is not well-constrained by the fit and it is broadly distributed, with $68\%$ of the probability lying between the glitch time calculated in \citet{palfreyman+2018} and 53.1 seconds before it. 
A strong correlation is also present between the glitch time, $t_g$, and the initial residual due to the magnetospheric slip.
In Figure~\ref{fig:fit}, we present the result of the fit, expressed as the median and the 16th and 84th percentiles, superimposed on the data. As we can remark based on this figure, the correlation is probably due to the fact that an anticipated glitch with a higher initial residual and a postponed glitch with a lower initial residual can fit the data equally well (see also \citealt{ashton+2019} about this). 

A tighter prior on the glitch time would allow for a better resolution on the probability distribution for the other parameters, for example $x_1$, which present a correlation of one of its peaks with the glitch time (see Figure~\ref{fig:9par}). 
The magnetospheric time $t_M$ presents two clear peaks, one at  6.4 s and one 2.6 s before the \citet{palfreyman+2018} glitch time. 
Unfortunately, the large uncertainty on $t_g$ does not allow us to conclude whether the magnetospheric change is before or after the triggering of the glitch.

\subsection{Timescales, overshoot parameter, and glitch size}

\begin{figure}
        \centering
        \includegraphics[width=\columnwidth]{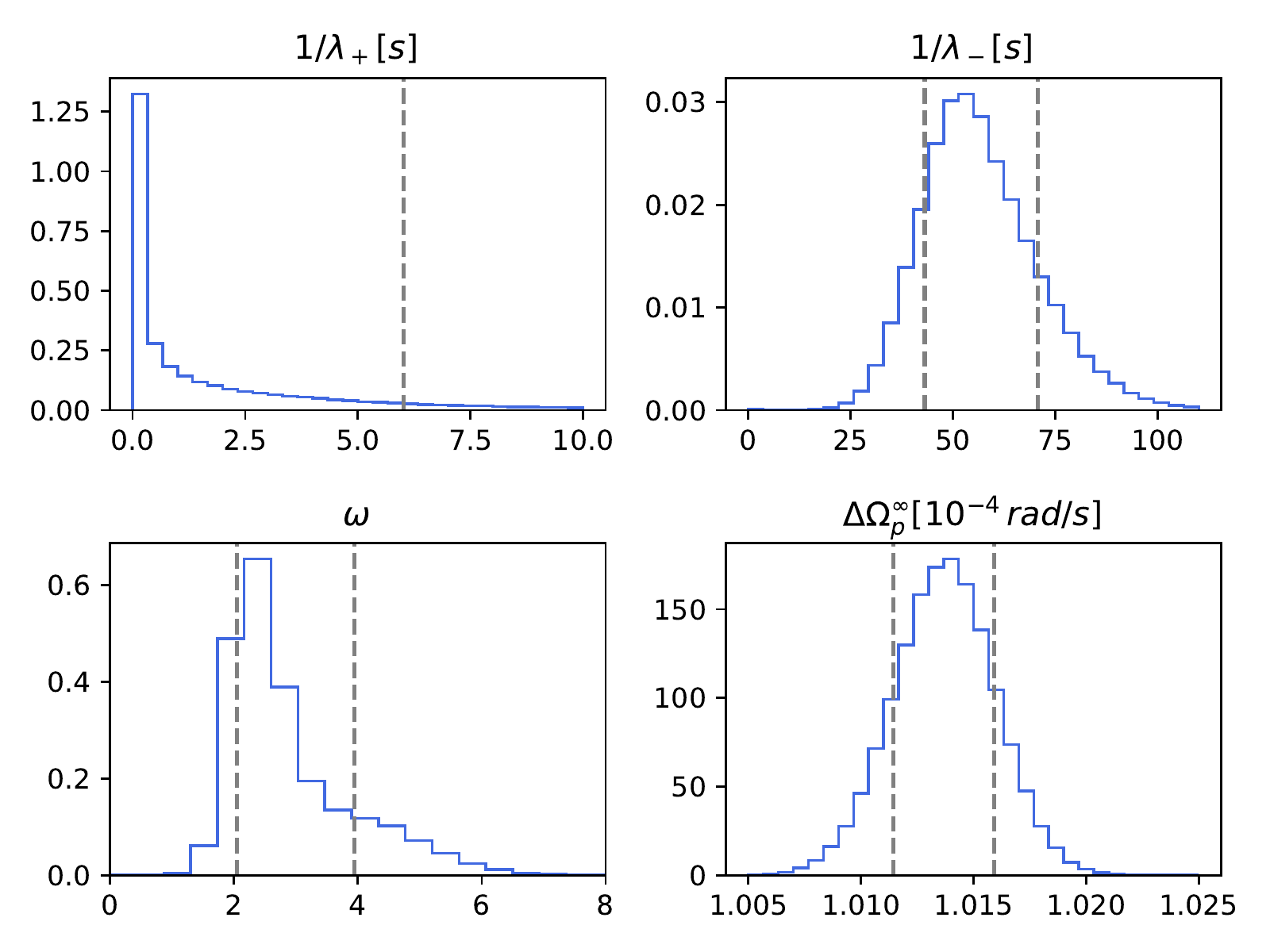}
        \caption{
        Probability distributions for the glitch rise timescale, $1/\lambda_+$, the relaxation timescale, $1/\lambda_-$, the overshoot parameter, $\omega,$ and the glitch size, $\Delta \Omega_{p}^\infty$. 
        For the glitch rise timescale, the 90th percentile is plotted, while for the other three quantities, the 16th and 84th percentiles are plotted.}
        \label{fig:quadratelli}
\end{figure}

The probability distributions for the rise timescale, $1/\lambda_+$, the relaxation timescale, $1/\lambda_-$, the overshoot parameter, $\omega,$ and the asymptotic glitch size, $\Delta \Omega_p^\infty$ , are given in Figure~\ref{fig:quadratelli}. 
The rise time, $1/\lambda_+,$ peaks close to 0s (the limit at which the rise is practically instantaneous) and $90\%$ of the distribution lies within 6.02s (cf. with Figure 2 of \citealt{ashton+2019}). 
This is a more stringent constraint with respect to the $\sim 12$s obtained in \citet{ashton+2019}, which is probably due to the different type of theoretical modelling underlying the fit (they used a single-timescale model to fit this parameter). 
The value obtained for the relaxation timescale is $1/\lambda_- = 55.07_{-11.99}^{+15.58}$ s: this value is also similar to that of previous glitches of the Vela; for example, the 2000 and the 2004 glitches \citep{dodson+2002, dodson+2007}.

Finally, the parameter $\omega$ obtained here has a value of $2.56_{-0.51}^{+1.38}$, which is a clear indication of the presence of an overshoot \citep{ashton+2019}, and the glitch size is $1.014 \pm 0.002 \cdot 10^{-4}$ rad/s, in good accordance with the previous estimate in Paper I.

\subsection{Comparison with a model with `active' and `passive' superfluid components}

For a better  comparison with the results in Paper I, we also performed a fit  by fixing $\Omega_{1p}^0=|\dot{\Omega}_\infty|/b_1$, the value corresponding  to the steady-state lag. 
In this way, we determine component 1 to be a `passive' one (in fact, a superfluid component that rotates with the steady state lag does not contribute to the angular momentum reservoir, which is the scenario considered in Paper I). 
In this case, we have to fit eight parameters instead of nine. 
We do not report the results here, as it yields fully compatible values for all the parameters shown in Figure \ref{fig:9par}. 
This establishes that the differences with respect to Paper I are mostly due to the difference in the fitting procedure and not to the assumption that component 1 is at a steady state at $t=0$ (i.e. it is `passive').
Moreover, the steady-state lag for the superfluid component 1, which is on the order of $10^{-8} - 10^{-9}$ rad/s, as calculated with the inferred values, is compatible to the results obtained here for the model with a free initial condition for $\Omega_{1p}^0$, again indicating a single reservoir. 

Regarding the Bayes factors, the eight-parameter model with $\Omega_{1p}^0$ fixed is only marginally preferable to that with a free initial condition, having a Bayes factor, $Z,$ of $\ln Z \approx 1.4$, which is too low to claim a strong preference between the two models \citep{kassraftery1995}.

Some considerations can be made for the other initial lag, $\Omega_{2p}^0$, which is distributed with a probability of the  $68\%$ in the range  $ 3 \times 10^{-4} \div   1 .1 \times 10^{-3}$ rad/s. 
In the years just before the glitch considered here, the Vela had undergone two glitches, as reported by the Jodrell Bank Glitch Catalogue\footnote{\url{ http://www.jb.man.ac.uk/pulsar/glitches.html}} \citep{espinoza+2011}: one in 2014, which is at least three orders of magnitude smaller than the one considered here, and one in 2013, which is the largest ever achieved in the Vela and of a comparable size with respect to that of 2016.
Starting from the equation for the conservation of angular momentum in system~\eqref{eq:3c}, we can notice that if we assume perfect pinning, ($\dot{\Omega}_2 = 0$) and $\dot{\Omega}_1 \approx \dot{\Omega}_p$, the angular velocity lag $\Omega_{2p}$ builds up at a constant rate,
\begin{equation}
    \dot{\Omega}_{2p} = - \dot{\Omega}_p = \frac{|\dot{\Omega}_\infty|}{1 - x_2} \, .
\end{equation}
Following \cite{montoli2020eos}, we estimate the lag accumulated before the 2016 glitch by assuming that the largest glitch (the one in 2013) has completely emptied the angular momentum reservoir (i.e. the lag between the components is null after the glitch). Furthermore, the 2014 glitch is so small that it is not expected to empty the accumulated reservoir substantially.
In this case, the expected angular velocity lag $\Omega_{2p}$ just before the 2016 glitch is
\begin{equation}
    \Omega_{2p} = \int_{t_{2013}}^{t_{2016}} \!\!\!\!dt\, \frac{|\dot{\Omega}_\infty|}{1 - x_2} \, 
    \gtrsim 
    (t_{2016}-t_{2013}) |\dot{\Omega}_\infty| \approx 0.01\,\text{rad/s}\,,
\end{equation}
which is one order of magnitude larger than what obtained from the fit.
This discrepancy can be interpreted in terms of vortex-creep: since a lag of $\sim0.01$ rad/s is expected by assuming perfect pinning of the superfluid 2 in the period between the 2013 and 2016 glitches, one possibility is that vortex creep is realised in place of perfect pinning, so that only the $10\%$ of the maximum achievable lag is actually stored. 

Finally, Figure~\ref{fig:9par} reveals the presence of a strong correlation between the moment of inertia fraction $x_2$ and the lag $\Omega_{2p}^0$: a different prior on the superfluid fraction, $x_2$, peaked or constrained to smaller values due, for example, to microphysical constraints, would give a smaller posterior value for it, thus yielding larger values of the initial lag.


\section{Physical interpretation of the fit}
\label{sec:physical}

We now discuss what information can be extracted from the fitted values of the phenomenological parameters, $x_i$ and $b_i$.
The physical interpretation of $x_i$ and $b_i$ is a slightly subtle matter due to the possible presence of entrainment between each superfluid component and the normal component (see Appendix \ref{app_entr}).

\subsection{Moments of inertia and mutual friction parameters}

To take into account the entrainment coupling, we interpret the lags between the two superfluid components and the normal one according to \eqref{pigna} and \eqref{bonzone}, so that a system like that of Equation~\eqref{eq:3c} still holds without the need to encode additional `entrainment torques'.
The downside is that the fractions of the moments of inertia, $x_i$ , contain a dependence on the entrainment parameter \citep{antonellipizzochero2017},
\begin{equation}
        x_i = \frac{8 \pi}{3 I} \int_0^R \mathrm{d}r\, r^4 \frac{\rho^i_n(r)}{1 - \epsilon^i_n(r)} = \frac{I_v^i}{I}
        \label{eq:xi}
,\end{equation}
where $I$ is the total moment of inertia, $R$ is the radius of the star, $\rho^i_n$  and $\epsilon^i_n$ are the mass density, and the entrainment parameter of the $i=1,2$ component. 
The parameter $I_v^i$ is the moment of inertia for the relative superfluid component corrected by entrainment, whereas the actual region that contributes to the integral is where $\rho^i_n>0$, see \eqref{Iv}. For zero entrainment, $I_v^i$ reduces to $I_n^i$.
We note that it is still possible to enforce relation \eqref{uno}, even in the presence of entrainment. As discussed in Appendix~\ref{app_entr}, this is because $x_p$ also changes due to entrainment (see \eqref{xp_ent}), in a way that keeps the value of total angular momentum \eqref{Ltotale} constant \citep[see also][]{antonelli+2018}.

Similarly, also the coupling parameters $b_i$, when expressed as spatial averages over some internal region of the star,  contain some entrainment correction (see \eqref{struzzo} and \eqref{bbb}),
\begin{equation}
        b_i \, = \, 2\Omega_p^0 \,\frac{8 \pi}{3 \, I_v^i} \int_0^R \mathrm{d}r\, r^4 \, \frac{\rho^i_n(r) \, \mathcal{B}_i(r)}{(1 - \epsilon^i_n(r)\, )^2} \, .
        \label{eq:bbbbbi}
\end{equation}
Here, $\mathcal{B}_i$ is the dimensionless mutual friction coefficient, usually expressed in terms of the drag-to-lift ratio $\mathcal{R}_i$ \citep[see e.g.][]{andersson+2006,sourie2020mnasL} as
\begin{equation}
        \mathcal{B}_i(r) \, = \, \frac{\mathcal{R}_i(r)}{1+\mathcal{R}_i(r)^2} \, .
        \label{eq:BBBBBi}
\end{equation}
Both $\mathcal{B}_i$ and  $\mathcal{R}_i$ are expected to have a spherical radial dependence as their value depends on the physical quantities in the stellar interior and on the particular mechanism that acts to dissipate energy at the microscopic scale of a vortex core. 

The coupling parameters, $b_1$ and $b_2$ , yield some information about the phenomena which cause the interaction between the superfluid component and the normal component. For the core superfluid, it is thought that electron scattering off magnetised vortices causes the drag between the superfluid and the normal component and then the subsequent exchange of angular momentum \citep{alpar84rapid}.
For the crustal superfluid, two different phenomena may occur, whether the relative velocity between the two components is small (phonon excitation, \citealt{jones1990}) or large (Kelvin waves excitation, \citealt{jones1992, epsteinbaym1992}). 
These two phenomena are believed to yield coupling parameters with rather different orders of magnitude. 

If we interpret the results obtained here for $b_1$ and $b_2$ as the coupling parameters for the core and the crustal superfluid (which seems unlikely given the posterior distribution of $x_2$, as discussed in the following subsection), respectively, then we can compare these results with the theoretical calculations done in the literature. 
From \eqref{eq:bbbbbi} it is immediate to obtain
\begin{equation}
        \begin{split}
        & \langle \, \mathcal{B}  \, \rangle_{\text{crust}} \, \approx \, 
        \frac{ \langle \, 1-\epsilon_n  \, \rangle_{\text{crust}} }{2 \, \Omega_p^0 } \, b_2   \, \approx \, 
        0.03\, b_2(\text{s}^{-1})
        \\
        & \langle \, \mathcal{B}  \, \rangle_{\text{core}} \, \approx \, 
        \frac{ \langle \, 1-\epsilon_n  \, \rangle_{\text{core}} }{2 \, \Omega_p^0 } \,  b_1   \, \approx \, 
        0.007\,  b_1(\text{s}^{-1}) \, 
        \end{split}
        \label{eq:Bcrustcore}
,\end{equation}
where the average values $\langle \, 1-\epsilon_n  \, \rangle_{\text{crust}} \approx 4$ and  $\langle \, 1-\epsilon_n  \, \rangle_{\text{core}} \approx 1$ have been taken from \cite{chamel2012} and \cite{chamelhaensel2006} respectively, while 
$\Omega_p^0 \approx 70\,\text{rad/s} $ has been employed.
Using the percentile values in Table \ref{tab:9par}, we obtain:
\begin{equation}
        \begin{split}
        & \langle \, \mathcal{B}  \, \rangle_{\text{crust}} \, \approx \, 
     2.4\times 10^{-3} \, -  \,  0.7
        \\
        & \langle \, \mathcal{B}  \, \rangle_{\text{core}} \, \approx \, 
        2.8\times 10^{-5} \, -  \,  6.3 \times 10^{-5} \, .
        \end{split}
        \label{eq:Bcrustcore2}
\end{equation}
However, if the crustal lattice is amorphous or contains a large number of defects, only weak entrainment is expected \citep{sauls+2020}, so we may use   $\langle \, 1-\epsilon_n  \, \rangle_{\text{crust}} \approx 1$ and obtain
\begin{equation}
         \langle \, \mathcal{B}  \, \rangle_{\text{crust}} \, \approx \, 
     5.6\times 10^{-4} \, -  \,  0.17
        \label{eq:Bcrusttttt}
.\end{equation}
The orders of magnitude of the coupling parameters calculated here are in good agreement with the most recent theoretical calculations for both the crust \citep{graber+2018} and the core superfluid \citep{andersson+2006}.
While this is a Newtonian model, a fully relativistic model would yield values for 
$\langle \, \mathcal{B}  \, \rangle_{\text{crust}}$  corrected by a factor of the order of $\approx 2$ \citep{sourieRise2017,gavassino+2020}. 

\subsection{Extension of the superfluid regions}

The fitted values for $x_{1,2}$ allow us some room for speculating on the spatial extension of the angular momentum reservoir. 
Similarly to Paper I, the results show that nearly the $x_1\approx 60 \%$ of the total moment of inertia refers to the component with a smaller initial lag (i.e. the component that before the glitch was likely to be only weakly pinned, so it did not develop a substantial lag).
On the other hand, we find $x_2\approx 15\%$ for the `strongly pinned' superfluid. 
This value is too large to be accommodated in the crust of the star alone, whatever the value of the entrainment in the crust, thus requiring that some of the reservoir superfluid should be located in the core of the star \citep{ho+2015, montoli2020eos}.

\begin{figure}
        \includegraphics[width=\columnwidth]{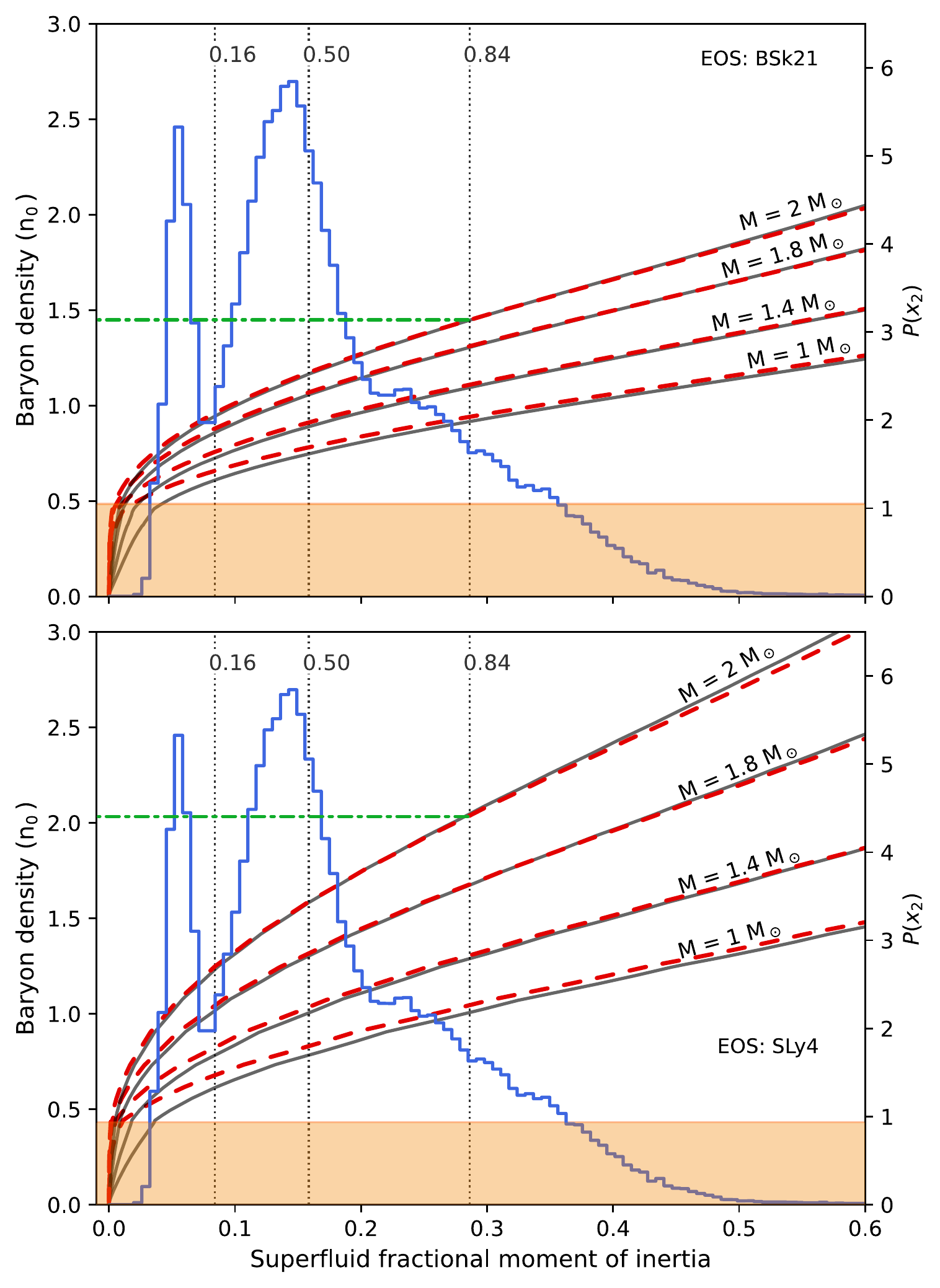}
        \caption{
    Comparison between the possible values of the superfluid moment of inertia fraction $I_v(n_B)/I$ and the possible values of $x_2$ according to its posterior distribution $P(x_2)$.
    We plot the points $(I_v(n_B)/I \, , \, n_B)$, where the baryon density $n_B$  is expressed in units of the nuclear saturation density $n_0=0.17\,$fm$^{-3}$. The inner-crust, i.e. $n_d<n_B \lesssim 0.5 n_0$, corresponds to the orange-shaded region.
        The upper panel refers to the BSk21 EoS \citep{goriely+2010}, the lower one to the SLy4 EoS  \citep{douchinhaensel2001}.
        The curves represent the points $(I_v(n_B)/I \, , \, n_B)$ when entrainment corrections are included (red-dashed) by using the values calculated by \cite{chamelhaensel2006} and \cite{chamel2012} and  when entrainment coupling is set to zero (gray-solid).
    The posterior $P(x_2)$ is superimposed as a background histogram, with the 16th, 50th and 84th percentiles shown with black dotted lines. 
    The two green dash-dotted lines indicate an upper limit (corresponding to the $84$th percentile) to the extension  of the superfluid 2 region if the mass of the Vela is $M=2 M_\odot$. 
        }
        \label{fig:xn}
\end{figure}

This can be seen in Figure~\ref{fig:xn}: here, we plot the moment of inertia fraction $I_v(n_B)/I$ of a spherical shell extending from a radius $R(n_B)$ to the radius $R(n_d)$,
\begin{equation}
\frac{I_v(n_B)}{I}  \, = \, 
\frac{8 \pi}{3 I} \int_{R(n_d)}^{R(n_B)} \mathrm{d}r\, r^4 \frac{\rho_n(r)}{1 - \epsilon_n(r)} \, ,
        \label{eq:procionebastardo}
\end{equation}
where $n_B$ is a generic baryon density such that $n_B>n_d$, while  $n_d$ is the drip-point density that defines the boundary between the inner and outer-crust.

We can try to match the theoretical value $I_v(n_B)/I$ with the fitted value of $x_2$. This would tells us that the superfluid 2 region extends between the densities $n_d$ and $n_B$. 
However, differently from what was done in Paper I,  the Bayesian fit does not provide a single value for $x_2$, but a posterior distribution (see Figure \ref{fig:9par}). 
For this reason, in Figure \ref{fig:xn}, along with $I_v(r)/I$, we also superimpose the posterior $P(x_2)$.  

The fraction $I_v(n_B)/I$  is calculated for different masses and two different unified equations of state (EoS), SLy4 \citep{douchinhaensel2001} and BSk21 \citep{goriely+2010}. 
We plot the cases with (red dashed lines) and without (grey solid lines) entrainment, where the coefficients $\epsilon_n$ for the core and the crust of the star are taken from \citet{chamelhaensel2006} and \citet{chamel2012}, respectively. 
Although $P(x_2)$ is doubly peaked, even the narrower peak on the left lies outside the crustal region for both the EoSs and for all the cases considered (with or without entrainment and for different masses). 
Moreover, this peak falls rapidly to zero for $x_2\lesssim 0.05$: it is for this reason that in all the cases considered, the value of $I_v(n_B)/I$ calculated at the crust-core interface lies in a region with very small or null values of $P(x_2)$, and well outside the 16-84 percentile region.

To check this result, we replicate the fit, but imposing the condition that  $x_1+x_2<0.05$ and keeping all the priors on the other parameters untouched. 
In this way, we limit the moment of inertia fraction to a portion that should coincide mostly with the crust of the star:
this value is an upper limit to the moment of inertia of the unbound neutrons in the crust when realistic equations of state are taken into account (see e.g. Figure 3 in \citealt{antonelli+2018}).

With the restriction $x_1+x_2<5\% $, we obtain non-physical posteriors for some of the parameters, in particular for the glitch rise time $t_g$, the initial residual $\Delta r_0$ and the magnetospheric time $t_M$. 
More importantly, since the nested sampling algorithm allows to estimate the evidence of the two models (the one with  $x_1+x_2<1$ and the one with  $x_1+x_2<5\%$), the natural logarithm of the Bayes factor between the two models is $\approx 5.6, $  favouring the model with $x_1+x_2<1$. 
A Bayes factor, $Z,$ such that $\log Z >5$ can be considered a strong evidence for a model with respect to another one \citep{kassraftery1995}. 
This test thus confirms the necessity of the inclusion of the superfluid in the core for the glitch process. 
We note that differently from the earlier results of  \cite{andersson+2012} and \cite{chamel2013}, 
the present result is independent of the presence of strong entrainment in the crust: this is because the imposed constraint, $x_1+x_2<5\%$ , can easily accommodate the fraction of the moment of inertia of the superfluid component in the whole crust either with or without entrainment corrections. 

Finally, considering the value of $x_2\approx 0.3$ at the 84th percentile as an upper limit to $I_v(r)/I$, from Figure \ref{fig:xn}, we can also conclude that the region relative to this superfluid component is the one extending from the drip point to $n_B \approx 1.5 n_0$ at most (for the BSk21 EoS and a star of $2 M_\odot$, as indicated by the horizontal dash-dotted line in the upper panel).
Similarly, we find that the region corresponding to component 2 extends, at most, up to  $n_B \approx 2 n_0$ if the Sly4 EoS is used.

%


\section{Conclusions}
\label{sec:conclusions}

Motivated by previous analyses of the 2016 Vela glitch \citep{ashton+2019}, we studied the minimal analytical model that is able to describe a pulsar glitch with overshoot, which requires three rigidly rotating components, where one is normal and two are superfluid. 


First, we improved the solution of the model presented in Paper I \citep[][]{pizzochero+2019}: we derived an analytic form for the time evolution of the observable component angular velocity (the calculations are in Appendix \ref{sec:solution}) and found the overshoot condition, dropping the assumption that the initial condition for one of the two superfluids is that of being at the steady state (we called such a component `passive' in Paper I). 
Moreover, we obtained the constraint on the fraction of the moment of inertia for the `slow'  superfluid component from \cite{sourie2020mnasL} in the present general formulation that includes superfluid entrainment; see \eqref{chebello2}.

We performed a Bayesian fit of the phenomenological parameters of the model using the data obtained by \citet{palfreyman+2018} for the 2016 Vela glitch: the basic form of the fitted function is identical to the one used by \citet{ashton+2019}, however, here, we write it in terms of the physical parameters of the system. 
The presence of a rise of the mean of the residuals close to the expected glitch time requires us to model this phenomenon. We decided to use a minimal  `magnetospheric slip' model, in which at a time, $t_M$  \textit{} -- which is a priori different from the glitch time ,$t_g$ -- the magnetosphere instantly decouples from and then recouples to the crust of the star, with a resulting apparent deceleration of its rotation. 
This model, while it is likely to be oversimplified, allows us to fit the data   and also  account  for this additional phenomenon that cannot be modeled with just a three-component model. 

Based on the fit, we estimated the coupling parameters between the superfluid components and the normal component, which have  orders of magnitude compatible to those obtained in theoretical calculations for the drag given by Kelvin wave excitation \citep{graber+2018} and electron scattering off magnetised vortices \citep{alpar84rapid,andersson+2006}. 
It has also been possible to obtain the fraction of the moment of inertia  of the two superfluid components of the model.
The marginalised posterior for the moment of inertia ratio of the superfluid, which acts as a main angular momentum  reservoir is rather broad. Nevertheless, it gives a clear indication that it is unlikely that the superfluid reservoir is limited to the crust of the star.

This claim has been double-checked by tightening the prior for the superfluid moment of inertia fractions to values that are similar to the crustal superfluid moment of inertia in a light star without entrainment. The evidence for this model is much smaller than that with the larger prior, confirming the unlikelihood of a superfluid reservoir limited to the crust of the star.
In contrast to what discussed in \cite{andersson+2012} and \cite{chamel2013}, the strength of this result is its independence from the entrainment parameter.

It has been possible to obtain the angular velocity lag between the pinned component and the normal component at the moment of the glitch, and it turned out to be an order of magnitude smaller with respect to the maximum lag achievable by Vela between the 2013 and the 2016 glitches. 
This can be interpreted in terms of the presence of vortex creep inside the star in the three years before the 2016 glitch, which turns out to be very efficient in dissipating the lag that could be built up in between glitches. 

The fit on the angular velocity of the star following the glitch allows us to calculate some other interesting quantities, such as an upper bound on the glitch rise timescale of $\sim 6\,$s and the following relaxation timescale (similar to that measured in other Vela glitches, such as \citealt{dodson+2002, dodson+2007}). 
The theoretical formalisation and the subsequent fit of the `overshoot parameter' of $\omega$ allow us to confirm the presence of an overshoot in the 2016 Vela pulsar glitch \citep{ashton+2019, pizzochero+2019}.  

Finally, we would like to stress the importance of a Bayesian approach in tackling this kind of problem: our previous knowledge of the Vela pulsar can be used in choosing the prior for the Bayesian inference. As this was the first pulse-to-pulse observation of a glitch, not much information can be inserted into the model. 
As more glitches of the Vela are recorded, however, more information on the particular parameters that we do not expect to change from glitch to glitch (such as the coupling parameters) can be gathered and used as priors for future observations. On the other hand, it  may also be the case that the analysis of a new glitch will give very different results for the moment of inertia fractions or the coupling parameters, indicating that the location of the superfluid regions that undergo unpinning depend on the past history of the star. 

\begin{acknowledgements}

Partial support comes from PHAROS, COST Action CA16214, and from INFN, the Italian Institute for Nuclear Physics. 
The authors would like to thank Gregory Ashton for fruitful discussions. 
Marco Antonelli acknowledges support from the Polish National Science Centre grant SONATA BIS 2015/18/E/ST9/00577, P.I.: B. Haskell.

\end{acknowledgements}


\appendix
\section{Solution to the system}
\label{sec:solution}

Here, we describe the procedure for solving the system in Equation \eqref{eq:3c}.
As a first step, we rewrite the component 3 system by performing a change of variables: it is convenient to use the superfluid angular velocities as measured in the frame of the normal component, $\Omega_{1p}$ and $\Omega_{2p}$.
Furthermore, it is convenient to directly integrate  the equation for $\Omega_p$ to find  that the angular velocity for the normal component with respect to the steady-state spin down solution is given by
\begin{equation}
\Delta \Omega_p(t) \,  := \,  \Omega_p(t) - \Omega_p^0 +  |\dot{\Omega}_\infty| \, t
\,  = \,
- \mathbf{x} \cdot (\mathbf{y}(t) - \mathbf{y}_0)\, , 
\label{eq:domP}
\end{equation}
where we have defined the vectors
\begin{equation}
\begin{split}
\mathbf{x} & = (x_1 \, , \, x_2)
\\
\mathbf{y} & = (\Omega_{1p} \, , \, \Omega_{2p}) \,  
\\
\mathbf{y}_0 & = (\Omega^0_{1p} \, , \, \Omega^0_{2p}) \, .
\end{split}
\label{cucciolotto}
\end{equation}
In this way, we only have to worry about the dynamics of the lag vector, $\mathbf{y}$, 
that must satisfy the matrix equation, 
\begin{equation}
\dot{\mathbf{y}} \, = \mathbf{a} - B \, \mathbf{y}    \, ,
\label{eq:pizzogallo}
\end{equation}
where
\begin{equation}
 \mathbf{a} = 
  \begin{bmatrix}
  \alpha \\
  \alpha
  \end{bmatrix}
 \, ,
 \qquad 
 B= 
 \begin{bmatrix}
 (1 - x_2) \beta_1 & x_2 \beta_2 \\
  x_1 \beta_1 & (1 - x_1) \beta_2
 \end{bmatrix}
\end{equation}
and 
\begin{align}
    &\alpha = |\dot{\Omega}_\infty| / (1 - x_1 - x_2)
    \\
    &\beta_i = b_i / (1 - x_1 - x_2) \qquad \text{for}\quad i=1,2\, .
\end{align}
The matrix $B$ has two eigenvalues $\lambda_{+}$ and $\lambda_{-}$,  given by 
\begin{equation}
        \lambda_{\pm} = \frac{1}{2}\left( T \pm \sqrt{T^2 - 4 D} \right),
        \label{eq:eigenvalues}
\end{equation}
where the positive parameters $T$ and $D$ represent the trace and the determinant of $B$, respectively. 
We call the respective eigenvectors, $\mathbf{e}_+$ and $\mathbf{e}_-$, defined up to a normalisation constant; their explicit form is not needed here. 

Using the fact that the parameters $b_i$ are positive and that the sum of the moment of inertia fractions $x_i$ cannot exceed unity, it is a straightforward exercise to prove that both the eigenvalues are always positive and in particular that $\lambda_+ > \lambda_- > 0$. 
Because of this positivity property, Equation \eqref{eq:pizzogallo} allows for a stable steady-state solution $\mathbf{y}(t)=\mathbf{y}_\infty$ that is constant in time: 
\begin{equation}
    \mathbf{y}_\infty = B^{-1} \mathbf{a} = ( \alpha / \beta_1 \, , \, \alpha / \beta_2) \, .
\end{equation}
This particular solution is an attractor for the dynamics of the lag vector $\mathbf{y}$:
the internal forces induced by dissipation (set by the parameters $b_i$) and the driving force (set by the parameter $|\dot{\Omega}_\infty| $) tend to balance out, killing off initial transients and settling the system into its typical behaviour described by $ \mathbf{y}_\infty$. 
Since we have two natural timescales (one short, $1/\lambda_{+}$, and one long, $1/\lambda_{-}$), we can conclude that the steady state is reached in the limit $t\gg 1/\lambda_{-}$.

The above property of the system allows to define the asymptotic amplitude of the glitch $\Delta\Omega_p^\infty$: 
we just have to take the limit $t\gg 1/\lambda_{-}$ in Equation \eqref{eq:domP} to obtain
\begin{equation}
    \Delta\Omega_p^\infty  = \mathbf{x}\cdot(\mathbf{y}_0 - \mathbf{y}_\infty ) \, .
\end{equation}
Instead of the lag vector $\mathbf{y}$, it is more convenient to consider the dynamics of the residual  with respect to the steady-state 
\begin{equation}
    \Delta\mathbf{y} = \mathbf{y}-\mathbf{y}_\infty \, ,
    \label{eq:residual_definition}
\end{equation}
that satisfies the linear equation, 
\begin{equation}
        \Delta \dot{\mathbf{y}} \, = \, -B \, \Delta \mathbf{y} \, .
        \label{eq:equation_linear}
\end{equation}
Decomposing $\bm{y}_0 - \bm{y}_\infty$ in the basis of the eigenvectors,
\begin{equation}
        \bm{y}_0 - \bm{y}_\infty \, = \, \delta y_- \mathbf{e}_- \, +\,  \delta  y_+ \mathbf{e}_+ \, ,
        \label{eq:decomp}
\end{equation}
the general solution of \eqref{eq:equation_linear} can be expressed as
\begin{equation}
\Delta \mathbf{y}(t) \, = \, e^{-t B} \, (\mathbf{y}_0-\mathbf{y}_\infty) \, = \, 
\sum_{j\, = \, +,-}  \mathbf{e}_j \,  \delta y_j \, e^{-t \lambda_j}    \, . 
\label{eq:sol_2_eq}
\end{equation}
%
%
Employing the  decomposition  \eqref{eq:decomp} and \eqref{eq:sol_2_eq} into \eqref{eq:domP}, it is easy to find
\begin{equation}
\Delta \Omega_p(t) 
\, = \, 
\Delta\Omega_p^\infty \left[ 1- \omega \, e^{-t \lambda_+} - (1-\omega)\,  e^{-t \lambda_-}  \right] \, ,
\label{eq:dOmp2}
\end{equation}
where we have defined  
\begin{equation}
\omega = \delta y_+ \, (\mathbf{x} \cdot \mathbf{e}_+) / \Delta\Omega_p^\infty \, .
\label{eq:www}
\end{equation}
Instead of using the eigenvectors, it is easier to find the value of $\omega$ in terms of the parameters of the system \eqref{eq:3c} by considering the value of the derivative of \eqref{eq:dOmp2} at $t=0:$
\begin{equation}
\label{omega}
\omega \, = \, 
\frac{ \Delta\dot{\Omega}_p(0) }{\Delta \Omega_p^\infty (\lambda_+ - \lambda_-) }
-
\frac{ \lambda_- }{ \lambda_+ - \lambda_- } \, .
\end{equation}
To write the general solution \eqref{eq:dOmp2} in terms of the basic parameters of the model, 
we need to know that
\begin{align}
\label{werty}
& \Delta {\Omega}^\infty_p \, = \, x_1 \left( \Omega_{1p}^0 - \frac{\alpha}{\beta_1} \right)
\, + \, x_2 \left( \Omega_{2p}^0 - \frac{\alpha}{\beta_2} \right)
\\
& \Delta \dot{\Omega}_p(0) \, = \, 
x_1 \, \beta_1 \, \Omega_{1p}^0 \, + \, x_2 \, \beta_2 \, \Omega_{2p}^0 \, - \, (x_1+x_2)\, \alpha 
\, ,
\label{qwer}
\end{align}
while the eigenvalues are given by
\begin{equation}
\begin{split}
        \lambda_\pm \, = \, & 
\frac{1}{2} \bigg[ 
\beta_1 (1-x_2) + \beta_2 (1-x_1) \pm    \\
& \pm \sqrt{ [ \beta_1 (1 - x_2) + \beta_2 (1 - x_1) ]^2 - 4 \beta_1 \beta_2 x_p }
\bigg]
\, .
\end{split}
\label{eq:lambdas}
\end{equation}
Finally, we observe that to obtain a positive glitch amplitude both $\Delta\Omega_p^\infty$ in \eqref{werty} and $\Delta\dot\Omega_p(0)$ in \eqref{qwer} should be positive. 
This constrains the initial lags and it is possible to show that this requirement is fulfilled for any possible value of $x_1$ and $x_2$ if
\begin{equation}
  |\dot{\Omega}_\infty| \, < \, \min_{i=1,2}[ \, b_i \, \Omega_{ip}^0  \, ] \, , 
  \label{piccione}
\end{equation}
which will be used in Appendix \ref{sec:constr}.

\section{Constraint on the moment of inertia of the slow component}
\label{sec:constr}

In Appendix \ref{sec:solution}, we presented the general solution to the three-component system, which extends the particular solution discussed in \cite{pizzochero+2019}. Building on this particular solution, \cite{sourie2020mnasL} recently proposed a simple formula to constrain the moment of inertia fraction of one of the superfluid component. 
It is worth to extend their treatment in view of the more general approach used here.

First, following \cite{sourie2020mnasL} we define the overshoot size $\Delta\Omega_{\rm over}$ as the maximum value touched during the spin-up phase. Using $\Delta\Omega_{\rm over} = \Delta\Omega_p( t_{\rm max} ) $ and Equation \eqref{eq:tmax}, we immediately obtain
\begin{equation}
 \frac{\Delta\Omega_{\rm over} }{ \Delta {\Omega}^\infty_p } 
= 
 1-
 \omega \left(\frac{\lambda_- \,  (\omega-1)}{\lambda_+ \omega}\right)^{\frac{\lambda_+}{\lambda_+ -\lambda_-}} 
 +
 (\omega-1) \left(\frac{\lambda_- \, (\omega-1)}{\lambda_+ \omega}\right)^{\frac{\lambda_-}{\lambda_+-\lambda_-}}  
 .
\label{eq:over}   
\end{equation}
This quantity depends on the phenomenological input parameters of the model (i.e. the $x_i$, $b_i$ and $|\dot{\Omega}_\infty| $) as well as on the initial condition $\Omega^0_{ip}$, for $i=1,2$. Up to this point, the role of the superfluid components 1 and 2 is symmetric (we do not assume Equation \eqref{break12} here) and all the formulas are invariant under the exchange of the two. 
However, let us assume that one of the two components, for example, that component 2 has a higher drag parameter with respect to the other, that is, $b_1 \ll b_2$ and 
\begin{equation}
a_{1/2} = b_1/b_2 = \beta_1/\beta_2 \ll 1 \, .
\label{asd}
\end{equation}
No further assumptions are needed on $x_1$ and $x_2$ (i.e. we do not need to specify which of the two components has higher inertia). This case is of physical interest (since we expect the nature and the strength of the friction mechanism to vary in different layers of the star) and allows to perform an expansion in the parameter $a_{1/2}$.

Under the assumption \eqref{asd}, the constraint \eqref{piccione} tells us that 
\begin{equation}
|\dot{\Omega}_\infty| \, < \, a_{1/2} \,  b_2 \, \Omega_{1p}^0
\qquad \text{and} \qquad
|\dot{\Omega}_\infty| \, < \,   b_2 \, \Omega_{2p}^0 \, .
\label{gufomarziano}
\end{equation}
Taking into account the above inequalities and inserting the expansions
\begin{align}
\label{wstar}
 & \omega = \omega^* + a_{1/2} \, \omega' +O(a_{1/2}^2)
 \\
 & \lambda_+  = \lambda_+^* + a_{1/2} \, \lambda_+' +O(a_{1/2}^2)
 \label{carciofo}
 \\
 & \lambda_-  = a_{1/2} \,  \lambda_-' +O(a_{1/2}^2) \, ,
\label{eq:expansion}   
\end{align}
into \eqref{eq:over}, it is possible to safely take the limit $a_{1/2}\ll 1$ to show that 
\begin{equation}
     \frac{\Delta\Omega_{\rm over} }{ \Delta {\Omega}^\infty_p } = \omega^* + O(a_{1/2} ) 
\end{equation}
and
\begin{equation}
     \frac{ \Delta\Omega_{\rm over} - \Delta {\Omega}^\infty_p}{ \Delta\Omega_{\rm over} } 
     = \frac{\omega^*-1}{\omega^*} + O(a_{1/2} ) 
     \, .
     \label{crispino}
\end{equation}
Thanks to \eqref{eq:lambdas}, we find that \eqref{carciofo} and \eqref{eq:expansion} state that:\ 
\begin{align}
 \lambda_+ &= \frac{b_2 \, (1-x_1)}{1-x_1-x_2} + \frac{a_{1/2} \, b_2 \,  x_1 \,  x_2 }{(1-x_1)(1-x_1-x_2)} +O(a_{1/2}^2)
 \\
 \lambda_- &= \frac{a_{1/2}  \, b_2}{1-x_1} +O(a_{1/2}^2) \, .
\end{align}
The lowest-order term $\omega^*$ in \eqref{wstar} can now be obtained by inserting the above equations into \eqref{eq:www}. 
Finally, the ratio in \eqref{crispino} turns out to be
\begin{equation}
\frac{ \Delta\Omega_{\rm over} - \Delta {\Omega}^\infty_p}{ \Delta\Omega_{\rm over} } 
     =  x_1 -  \frac{x_1(1-x_1) (b_1 \Omega_{2p}^0  - |\dot{\Omega}_\infty|) }{ b_1 \, x_2 \, \Omega_{2p}^0 } + O(a_{1/2} ) 
     \, ,
     \label{ciaspola1}
\end{equation}
or, equivalently, 
\begin{equation}
\frac{ \Delta\Omega_{\rm over} - \Delta {\Omega}^\infty_p}{ \Delta\Omega_{\rm over} } 
     =  x_1 - \frac{(1-x_1) (\Delta\Omega_p^\infty- x_2\Omega_{2p}^0)}{ x_2\, \Omega_{2p}^0} + O(a_{1/2} ) 
     \, .
     \label{ciaspola2}
\end{equation}
Equation \eqref{gufomarziano} tells us that the second term in the right hand side is always negative, such that to the lowest order in $a_{1/2}$, the detection of an overshoot allows to constrain the fractional moment of inertia of the `slow' component (in this case $x_1$) as
\begin{equation}
 x_1 \, > \, \frac{ \Delta\Omega_{\rm over} - \Delta {\Omega}^\infty_p}{ \Delta\Omega_{\rm over} }  
 \qquad \text{for} \quad  b_1 \ll b_2 \, .
     \label{chebello}
\end{equation}
This is in complete accordance with Equation (12) of \cite{sourie2020mnasL}. In Appendix \ref{app_entr}, we describe how to interpret $x_1$ when there is superfluid entrainment between the components, see Equation \eqref{chebello2}.

\section{Including entrainment}
\label{app_entr} 

It is straightforward to include the entrainment effect into our system of equations, provided that a convenient choice of the dynamical variables is made. 
In fact,  using the `superfluid momentum' of $\mathbf{p}_n$ instead of the `neutron velocity' of $\mathbf{v}_n$ naturally leads to a redefinition of the phenomenological parameters of the hydrodynamic model (here, $x_i$ and $b_i$ for $i=1,2$), but the form of the dynamical equations remains unchanged \citep{antonellipizzochero2017}. 
Originally, the argument was presented in the very special case of straight and rigid vortex lines in a Newtonian context, but it can be generalised to the case of `slack' vortices and of different superfluid domains, as well as to take into account for general relativistic corrections in the slow-rotation approximation \citep{antonelli+2018,gavassino+2020}. 

Differently from Paper I, here we derive Equation \eqref{eq:3c} starting from a local and fluid model. Hence, the present discussion is analogous to the one made by \cite{sidery_glitch_2010} and differs from it only with regard to the choice of variables, nonetheless, it is quite convenient in the present framework where we have to deal with three different components. The results of this section can be immediately extended to a generic number of non-overlapping superfluid components.

Locally, the  momentum per particle $\mathbf{p}_n$ of the superfluid neutrons is a linear combination of the
neutron velocity $\mathbf{v}_n$ and of the velocity of the normal component $\mathbf{v}_p$ (which is a mixture of all the charged species and we assume it to be rigid),
\begin{equation}
    \mathbf{p}_n/m_n = (1-\epsilon_n)\mathbf{v}_n + \epsilon_n \mathbf{v}_p \, ,
\end{equation}
where $m_n$ is the neutron mass and $\epsilon_n$ is the entrainment parameter \citep{haskellsedrakian2017,chamel_super}.

If we have two different (non-overlapping) superfluid regions and the motion is circular, the above equation suggests to define two additional angular velocities $\Omega_v^i$ as
\begin{equation}
    \Omega_v^i = (1-\epsilon_n^i)\Omega_n^i + \epsilon_n^i \Omega_p \, , \qquad \text{for }\quad i=1,2
    \label{pigna}
,\end{equation}
where $\Omega_p$ is the observable angular velocity of the normal $p$-component while $\Omega_n^i$ is the angular velocity of the neutrons in the region $i=1,2$.
Working with the $\Omega_v^i$ is convenient because, due to the Feynman-Onsager relation, they are a direct measure of the number of vortices in a certain superfluid region. 
Hence, the $\Omega_v^i$ cannot change as long as the number of vortices is conserved. 
This defines the form of the equations of motion at a certain location $\mathbf{x}$ inside the star \citep{antonellipizzochero2017},
\begin{equation}
    \partial_t \Omega^i_v(t,\mathbf{x}) \, \approx \, -2 \Omega^i_v(t,\mathbf{x}) \frac{\mathcal{R}_i}{1+\mathcal{R}_i^2} (\Omega^i_n(t,\mathbf{x}) - \Omega_p(t) )
    \label{eq:boh}
,\end{equation}
where $\mathcal{R}_i$ is the drag-to-lift ratio that appears in the vortex-mediated mutual friction force between the superfluid and normal components \citep{andersson+2006}. In Equation~\eqref{eq:boh}, we dropped a term $\partial_x \Omega_{v}^i$, absent for rigid rotation, hence the $\approx$ symbol.
With the aid of \eqref{pigna}, the above equation states:\ 
\begin{equation}
    \partial_t \Omega^i_v(t,\mathbf{x}) \, \approx \, - B_i(r) \, (\Omega^i_v(t,\mathbf{x}) - \Omega_p(t) ) \, ,
    \label{batufolo}
\end{equation}
where the coefficient $B_i(r)$ depends on the local values $\mathcal{R}_i(r) $ and $\epsilon^i_n(r)$ at a certain radius $r$ inside the star (we assume spherical stratification), namely,
\begin{equation}
    B_i(r)  =  \frac{2 \, \Omega^i_v}{1-\epsilon^i_n} \frac{\mathcal{R}_i}{1+\mathcal{R}_i^2} \approx
         \frac{2 \, \Omega_p}{1-\epsilon^i_n} \frac{\mathcal{R}_i}{1+\mathcal{R}_i^2} \, ,
         \label{struzzo}
\end{equation}
where $\Omega_v^i \approx \Omega_p$ because the lags are small.
When the variables $\Omega^i_v$ are used, the total angular momentum of the star $L_{\rm tot}$ is given by 
\begin{equation}
L_{\rm tot} = (I - I_v^1 - I_v^2) \Omega_p + I_v^1 \langle \Omega_v^1 \rangle_1 +  I_v^2 \langle \Omega_v^2 \rangle_2 \, ,
\label{Ltotale}
\end{equation}
where $I$ is the total moment of inertia and\footnote{
    Since the integration is over the $i$-region, and the two superfluid regions do not overlap, we can drop the unnecessary $i$ labels on the density and on the entrainment parameter. We do the same in \eqref{averagef}.
}
\begin{equation}
I_v^i \, = \, \frac{8 \pi}{3} \int_{i} dr \, r^4 \frac{ \rho_n(r) }{1-\epsilon_n(r)}   
\label{Iv}
\end{equation}
is a rescaled moment of inertia for the superfluid component (the integration extends over the region $i$ and $\rho_n(r)$ is the density of unbounded neutrons). 
Using standard spherical coordinates where $\theta$ is the colatitude, the parameters $I_v^i$ play the role of  normalisation factors for the averages of functions over the $i$-region,
\begin{equation}
\langle \, f \, \rangle_i  \, = \, 
\frac{1}{I_v^i} \int_{i} d^3x \,f(\mathbf{x}) \, \frac{(\sin\theta \, r)^2 \rho_n(r) }{1-\epsilon_n(r)} \, .  \label{averagef}
\end{equation}
We now take the spatial average of Equation \eqref{batufolo},
\begin{equation}
    \langle  \dot{\Omega}^i_v\rangle_i \, \approx \, - \langle B_i (\Omega^i_v  - \Omega_p  ) \rangle_i 
    \approx \, - \langle B_i \rangle_i \, \langle  \Omega^i_v  - \Omega_p   \rangle_i \, . 
        \label{batufolo2}
\end{equation}
Clearly, the last step is not rigorous but neglecting possible correlations between the local value of $B_i$ and the spatial fluctuations of the lag $\Omega^i_v  - \Omega_p$ is the price we have to pay to obtain a rigid model from a fluid one.
Finally, the only effect of the spin-down torque is to transport the angular momentum to infinity, so it can be introduced as 
\begin{equation}
\dot{L}_{\rm tot} \, = \, 
(I- I_v^1- I_v^2) \dot \Omega_p + I_v^1 \langle \dot \Omega_v^1 \rangle_1 +  I_v^2 \langle \dot \Omega_v^2 \rangle_2
\, = \,  - I |\dot\Omega_\infty| \, .
        \label{castoro}
\end{equation}
Equations \eqref{batufolo2} and \eqref{castoro} are identical to the system in \eqref{eq:3c}, provided that we make the following identifications:
\begin{align}
& x_i \, = \, I_v^i \, / \, I
\\
& x_p \, = \,  (I- I_v^1- I_v^2) / I = 1- x_1-x_2
\label{xp_ent}
\\
& b_i  \, = \,  \langle \, B_i \, \rangle_i
\label{bbb}
\\
&  \Omega_i \, = \, \langle \,\Omega_v^i  \, \rangle_i \, .
\label{identificazioni}
\end{align}
Similarly, the lag vector $\mathbf{y}$ in \eqref{cucciolotto} should be interpreted as
\begin{equation}
    \mathbf{y} \,= \, (\,  \Omega_v^1-\Omega_p   \, , \, \Omega_v^2-\Omega_p \, )\, .
    \label{bonzone}
\end{equation}
Including all the entrainment corrections into the phenomenological parameters has the advantage that the dynamical equations do not change, so that no new calculations are needed to find the solution of the system. Hence, the generalisation of the formula of \cite{sourie2020mnasL} to the entrained case is still given by~\eqref{chebello},~where  
\begin{equation}
 x_1 \, = \frac{I_v^1}{I}\, > \, \frac{ \Delta\Omega_{\rm over} - \Delta {\Omega}^\infty_p}{ \Delta\Omega_{\rm over} }  
\qquad \text{for} \quad  \langle \, B_1 \, \rangle_1 \ll  \langle \, B_2 \, \rangle_2 \, .
 \label{chebello2}
\end{equation}
\phantom{.}


\bibliographystyle{aa}
\bibliography{biblio}

\end{document}